\documentclass[10pt,conference]{IEEEtran}

\makeatletter
\newcommand*{\inlineequation}[2][]{%
  \begingroup
    \refstepcounter{equation}%
    \ifx\\#1\\%
    \else
      \label{#1}%
    \fi
    \relpenalty=10000 %
    \binoppenalty=10000 %
    \ensuremath{%
      #2%
    }%
    ~\@eqnnum
  \endgroup
}
\makeatother

\IEEEoverridecommandlockouts
\usepackage{cite}
\usepackage{amsmath,amssymb,amsfonts}
\usepackage[linesnumbered,ruled,vlined]{algorithm2e}

\usepackage{graphicx}
\usepackage{textcomp}
\usepackage[svgnames]{xcolor}
\usepackage{url}
\usepackage{xspace}
\usepackage{booktabs}
\usepackage{multirow}
\usepackage{lipsum}
\usepackage{hyperref}
\hypersetup{
    colorlinks = true,
    citecolor = {darkblue},
    urlcolor = {darkblue},
    linkcolor = {darkblue}
}

\definecolor{dimgray}{rgb}{0.41, 0.41, 0.41}
\definecolor{brickred}{rgb}{0.8, 0.25, 0.33}
\definecolor{cadmiumgreen}{rgb}{0.0, 0.42, 0.24}
\definecolor{dkgreen}{rgb}{0,0.6,0}
\definecolor{mauve}{rgb}{0.58,0,0.82}
\definecolor{gray}{rgb}{0.4,0.4,0.4}
\definecolor{darkblue}{rgb}{0.0,0.0,0.6}
\definecolor{lightblue}{rgb}{0.0,0.0,0.9}
\definecolor{cyan}{rgb}{0.0,0.6,0.6}
\definecolor{darkred}{rgb}{0.6,0.0,0.0}

\usepackage{listings}
\lstdefinelanguage{diff}{
  morecomment=[f][\color{dimgray}]{@@},
  morecomment=[f][\color{cadmiumgreen}]{+\ },
  morecomment=[f][\color{brickred}]{-\ },
}

\def\name{\textsc{Fonte}\xspace}
\def\product{SAP HANA\xspace}

\def\BibTeX{{\rm B\kern-.05em{\sc i\kern-.025em b}\kern-.08em
    T\kern-.1667em\lower.7ex\hbox{E}\kern-.125emX}}
\begin{document}

\title{\name: Finding Bug Inducing Commits\\from Failures}

\author{
\IEEEauthorblockN{Gabin An}
\IEEEauthorblockA{
\textit{School of Computing, KAIST}\\
Daejeon, Republic of Korea \\
agb94@kaist.ac.kr}
\and
\IEEEauthorblockN{Jingun Hong}
\IEEEauthorblockA{
\textit{SAP Labs Korea}\\
Seoul, Republic of Korea \\
jingun.hong@sap.com}
\and
\IEEEauthorblockN{Naryeong Kim}
\IEEEauthorblockA{
\textit{School of Computing, KAIST}\\
Daejeon, Republic of Korea \\
kimnal1234@kaist.ac.kr}
\and
\IEEEauthorblockN{Shin Yoo}
\IEEEauthorblockA{
\textit{School of Computing, KAIST}\\
Daejeon, Republic of Korea \\
shin.yoo@kaist.ac.kr}
}

\maketitle

\begin{abstract}
A Bug Inducing Commit (BIC) is a commit that introduces a software bug into
the codebase. Knowing the relevant BIC for a given bug can provide valuable
information for debugging as well as bug triaging. However, existing BIC
identification techniques are either too expensive (because they require the
failing tests to be executed against previous versions for bisection) or
inapplicable at the debugging time (because they require post hoc artefacts
such as bug reports or bug fixes). We propose \name, an efficient and accurate
BIC identification technique that only requires test coverage. \name combines
Fault Localisation (FL) with BIC identification and ranks commits based on the
suspiciousness of the code elements that they modified. \name reduces
the search space of BICs using failure coverage as well as a filter that
detects commits that are merely style changes. Our empirical evaluation using 130 real-world BICs shows that \name significantly
outperforms state-of-the-art BIC identification techniques based on Information
Retrieval as well as neural code embedding models, achieving at least 39\% higher MRR.
We also report that the ranking scores produced by \name can be used to perform
weighted bisection, further reducing the cost of BIC identification. Finally,
we apply \name to a large-scale industry project with over 10M lines of code,
and show that it can rank the actual BIC within the top five commits for 87\%
of the studied real batch-testing failures, and save the BIC inspection cost by
32\% on average.

\end{abstract}


\section{Introduction}

A Bug Inducing Commit (BIC)~\cite{liwerski2005} refers to a commit that
introduces buggy source code into the program. Accurate identification of BICs
can have many benefits. Existing work has shown that simply reverting BICs may
suffice for bug fixes~\cite{Wu2017, Wen2020}, while the knowledge of BICs can
aid manual debugging by developers~\cite{Wen2019}. Considering the finding that
78\% of bugs are fixed by those who introduced them in the first
place~\cite{Wen2016}, the knowledge of BICs can help effective assignment of a
newly revealed bug to the right team or developers~\cite{Murali2021}. Finally,
for software engineering researchers, a BIC dataset can be used to study how
bugs are created~\cite{RodrguezPrez2020,liwerski2005}, eventually resulting in
better software defect prediction techniques~\cite{Catal2011}.

Many approaches have been proposed to identify BICs: existing approaches can be
categorised into three groups. The first group is bisection, i.e., a binary
search on the commit history that checks whether each snapshot in the commit
history is buggy or not~\cite{gitbisect}. The actual inspection can be done either
manually, or automatically by executing the bug-revealing test cases. While the
binary search itself is efficient, the inspection cost required for each
snapshot can render bisection impractical. For example, with large-scale
software projects, the cost of simply building and executing test cases for a
specific snapshot can be significantly high.

While bisection depends on explicitly checking each snapshot, other approaches
are static, i.e., they only concern commit histories or bug reports. The second
group is represented by SZZ~\cite{liwerski2005} and its variants. Given a Bug
Fixing Commit (BFC), SZZ essentially seeks to identify a set of commits that
last modified each element of BFC. However, SZZ-like approaches require BFCs as input,
which are only available when the bug has already been patched. Consequently,
these techniques are not applicable at the debugging time. The third group is
IR-based BIC identification~\cite{Wen2016,Bhagwan2018}, which
reformulates BIC identification as Information Retrieval (IR)
where the bug report is the query, and the commits are the documents.
Given a query, i.e., the bug report
that contains various information about the failure in question, the BIC is
likely to be the commit that is the most lexically similar to the query. While
IR-based approaches do not incur the cost of their dynamic counterparts (e.g.,
compilation and test execution), they cannot be applied if a bug
report for the latest failure is not available yet, or if it does but its
quality is too low.

In this paper, we aim to propose a BIC identification technique that is
accurate, efficient, and available at debugging time. Intuitively, our technique,
\name\footnote{\name is an Italian word meaning ``source'' or ``origin''.}, distributes the suspiciousness computed by a Fault Localisation (FL)
technique for the current bug to commits in the development history, expanding
the dimensions of FL techniques from the location within the codebase (i.e., spatial) to the history of the codebase (i.e., temporal). \name starts by collecting the test coverage at the time of failure and
computing suspiciousness scores for code elements using a FL
technique~\cite{Wong2016}. Subsequently, \name traces back the commits that are
relevant to the code covered by failing tests~\cite{An2021}. Commits that are mere style
changes are filtered out based on  Abstract Syntax Tree (AST) level comparisons. Finally, the remaining
commits are ranked according to the suspiciousness of the current code that is
modified by each candidate commit.
Compared to bisection, \name does not require inspection of each snapshot it
considers and therefore more efficient. Instead, \name uses the failure
coverage to improve accuracy. It also does not require bug reports or bug
fixes, and therefore can be applied at debugging time once a test failure is
observed.

We evaluate \name with a benchmark of 130 real-world bugs: 67 from an existing
BIC dataset, and 63 that are manually curated by us. The results show that the
ranking of BICs produced by \name achieves 242\% of MRR compared to the
random baseline. Furthermore, exploiting the fact that \name assigns scores to
each candidate BIC, we also propose a weighted bisection method that leverages
the commit scores during the search. Weighted bisection combined with \name can
save the number of bisection iterations for 98\% of the cases. Since \name does
not require any manual human effort, it can be easily incorporated into CI
pipelines to provide developers with candidate BICs when reporting test
failures.

The contributions of this paper are summarised as follows:

\begin{itemize}
\item We present \name, a BIC identification technique that only requires
information from the failed test execution and the commit history. Since it
does not require bug patches or bug reports, \name can be used to aid debugging
by providing developers with the relevant BICs, once a test failure is observed.
\name builds upon our previous work~\cite{An2021} by actually quantifying the suspiciousness of commits instead of simply reducing search space.

\item We evaluate \name with 130 real-world bugs and show that \name can
accurately rank BIC candidates: it achieves an MRR of 0.528, which outperforms a state-of-the-art IR-based BIC identification
technique by 39\%.

\item We introduce weighted bisection that uses the scores assigned to
candidate BICs by \name. Weighted bisection can reduce the required number of
iterations for about 98\% of studied bugs when compared to the standard
bisection algorithm applied to the entire commit history.

\item We apply \name to the batch testing scenario of large-scale industry software. It achieves 547\% of MRR compared to the random baseline and can reduce the bisection iterations in 78\% of cases. 

\item We release a new BIC benchmark dataset for 130 Defects4J version 2.0 bugs. \name is publicly available at \url{https://github.com/coinse/fonte}, along with artefacts of the empirical
evaluation in this paper.
\end{itemize}

The remainder of the paper is structured as follows. Section~\ref{sec:background} explains the research context of this paper and defines the basic notations. Section~\ref{sec:methodology} and \ref{sec:weighted_bisection} propose \name and the novel weighted bisection method, respectively. Section~\ref{sec:eval_setup} describes the empirical evaluation settings for \name along with the research questions, and Section~\ref{sec:results} presents the results. Section~\ref{sec:industry} shows the application results of \name to the batch testing scenario in industry software. Section~\ref{sec:threats} addresses the threats to validity, and Section~\ref{sec:related_work} covers the related work of \name. Finally, Section~\ref{sec:conclusion} concludes.

\section{Background}
\label{sec:background}

This section provides the background of this paper.

\subsection{Research Context}
Debugging is usually initiated by observing a failure that reveals a bug in the
program. Even when a field failure is reported by users, the debugging
activities typically start with reproducing the field failure~\cite{Artzi, Jin2012, Zimmermann2010}: this is
because failure-triggering test cases are essential to confirm whether the bug
is fixed or not. Once observed, the test failure goes through the bug triage
phase to be assigned to a developer or a team, who will analyse the buggy
behaviour and produce a patch.
Knowing the BIC responsible for the observed failure can not only contribute to
more efficient bug triage~\cite{Murali2021} but also help developers better
understand the context of the buggy behaviour~\cite{Wen2016}. To identify BICs as soon as the bug is detected, we cannot rely on any information that is produced later in the debugging process, such as bug fixes or bug reports.

While some BIC identification techniques~\cite{Bhagwan2018,Murali2021} are
based on the information from failures, they only use the stack traces or the
exception messages, which may only be indirectly linked to the contents of
actual BICs. Given that commits are directly coupled to specific locations
in the source code, we focus on the actual coverage of the failing tests as the
main source of information. Our previous work shows
that the coverage of failing test
executions (i.e., \emph{failure coverage}) can reduce the BIC search space
very effectively~\cite{An2021}: simply by filtering out any commit that is not
related to the evolution of code elements covered by the failing tests, the
search spaces of BICs for 703 bugs in Defects4J~\cite{Just2014} were
reduced to 12.4\% of their original size on average. The high reduction rate suggests that failure coverage has the potential to provide the basis for a BIC identification technique available at the debugging time.

In this work, we aim to accurately locate the BIC using only the information
that is available at the debugging time, just after the observation of test
failure. We build upon the previous technique of BIC search space
reduction~\cite{An2021}, and present a technique that can accurately measure
the relevance of each commit in the reduced search space to the failure.
Intuitively, our approach distributes the code level suspiciousness measured in
the current buggy version to the past commits.

\subsection{Basic Notations}
\label{sec:background:notation}

Let us define the following properties of a program $P$:
\begin{itemize}

  \item A set of commits $C = \{c_1, c_2, \ldots\}$ made to $P$
  \item A set of code elements $E = \{e_1, e_2, \ldots\}$ of $P$, such as
  statements or methods

  \item A set of test cases $T = \{t_1, t_2, \ldots\}$ where $T_F \subseteq T$ is a set of failing test cases
\end{itemize}

We assume that the bug responsible to the failing tests
resides in the source code, i.e., some elements in $E$ cause the failure of $T_F$. We also define the following relations
on sets $C$, $T$, and $E$:

\begin{itemize}
  \item A relation $\mathsf{Cover} \subseteq T \times E$ defines the relation between test cases and code elements. For every $t \in T$ and $e \in E$, $(t, e) \in \mathsf{Cover}$ if and only if the test $t$ covers $e$ during the execution.
  \item A relation $\mathsf{Evolve} \subseteq C \times E$ defines the relation between commits and code elements. For every $c \in C$ and $e \in E$, $(c, e) \in \mathsf{Evolve}$ if and only if the commit $c$ is in the change history of the code element $e$.
\end{itemize}

As our ultimate goal is to find the BIC in $C$, we aim to design a scoring function $s \colon C \to \mathbb{R}$ that gives higher scores to commits that have a higher probability of being the BIC.

\section{\name: Automated BIC Identification via Dynamic, Syntactic, and Historical Analysis}
\label{sec:methodology}

This paper presents \name, a technique to automatically identify the BIC, based
on the assumption that \emph{a commit is more likely to be a bug inducing
commit if it introduced or modified a code element that is more relevant to
the observed failure}. The key idea behind \name is that the relevancy of the code elements to the observed failures can be quantified using existing FL techniques~\cite{Wong2016}, such as SBFL. \name distributes the code-level suspiciousness,
computed by an FL technique for the observed failures, to commits in the development history. Fig.~\ref{fig:overview} illustrates the three stages of
\name, which are described below:

\begin{figure}[t]
  \centerline{\includegraphics[width=0.85\linewidth]{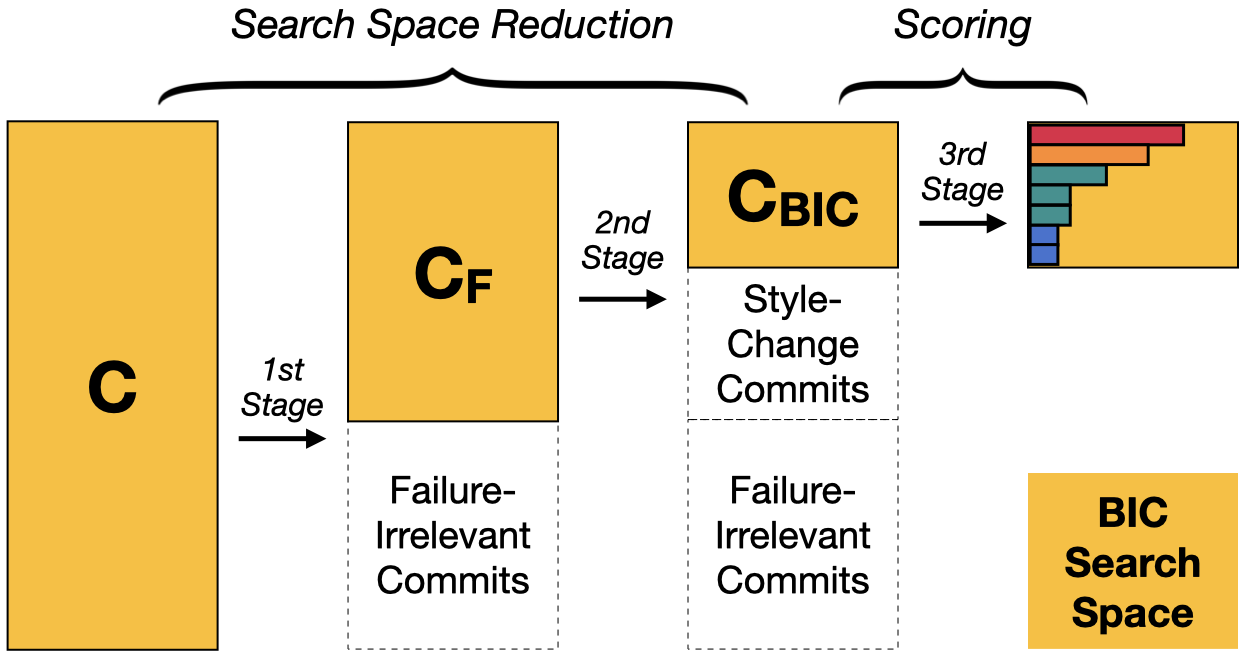}}
  \caption{Overview of \name}
  \label{fig:overview}
\end{figure}

\begin{enumerate}
\item \name identifies all suspicious code elements using the coverage of failing test cases and discards the commits that are irrelevant to the code elements from the BIC search space~\cite{An2021}.
\item \name additionally filters out the commits that contain only style changes to the suspicious files using AST level comparisons.
\item \name assigns scores to the remaining commits in the search space using the FL scores and evolution history of the suspicious code elements.
\end{enumerate}

The rest of this section describes each stage in more detail.

\subsection{Stage 1: Filtering Out Failure-Irrelevant Commits}
\label{sec:methodology:stage1}

Using the notations defined in Section~\ref{sec:background:notation}, we can
represent the failure-coverage-based BIC search space reduction~\cite{An2021} as follows. First, let $E_{F} \subseteq E$ denote the set
of all code elements that are covered by the failing test cases:

\begin{equation}
  \label{eq:E_susp}
E_{F} = \bigcup_{t \in T_F}\{e \in E|(t, e) \in \mathsf{Cover}\}
\end{equation}

Subsequently, we obtain $C_{F} \subseteq C$, a set of commits that are involved in the evolution of at least one code element in $E_{F}$:
\begin{equation}
  \label{eq:C_susp}
C_{F} = \bigcup_{e \in E_{F}}\{c \in C|(c, e) \in \mathsf{Evolve}\}
\end{equation}

Then, all commits not contained in $C_{F}$ can be discarded from our BIC
search space because the changes introduced by those commits are not related to
any code element executed by failing executions. Consequently, the BIC search
space is reduced from $C$ to $C_{F}$.

\subsection{Stage 2: Filtering Out Style Change Commits}
\label{sec:methodology:stage2}

\begin{figure}[t]
\begin{lstlisting}[language=diff]
@@ -74,7 +74,7 @@ import org.apache.commons.lang.exception.NestableRuntimeException;
  * @author Phil Steitz
  * @author Pete Gieser
  * @since 2.0
- * @version $Id: StringEscapeUtils.java,v 1.26 2003/09/07 14:32:34 psteitz Exp $
+ * @version $Id: StringEscapeUtils.java,v 1.27 2003/09/13 03:23:24 psteitz Exp $
  */
 public class StringEscapeUtils {
 
@@ -242,7 +242,9 @@ public class StringEscapeUtils {
             } else {
                 switch (ch) {
                     case '\'':
-                        if (escapeSingleQuote) out.write('\\');
+                        if (escapeSingleQuote) {
+                          out.write('\\');
+                        }
                         out.write('\'');
                         break;
                     case '"':
\end{lstlisting}
  \caption{Changes by the commit \texttt{5814f50} in Defects4J \texttt{Lang-46}}
  \label{fig:Lang-46b-5814f50}
\end{figure}

The reduced set of candidate BICs, $C_{F}$, may still contain \emph{style
change commits}, i.e., commits that do not introduce any semantic change to the
suspicious code elements. These commits can be further excluded from the BIC
search space, as they cannot have altered the functional behaviour of
the relevant code elements~\cite{Kim2006}. An example of such a commit is shown in Fig.~\ref{fig:Lang-46b-5814f50}, which modifies the comments and encloses the single statement in the \texttt{if} block.

We use the AST level comparison~\cite{Falleri2014} to identify whether a given
commit $c \in C_{F}$ is a style change commit or not. First, we identify the
set of source files, $S$, that are modified by the commit $c$ and covered by the
failing test cases. Formally, any file in $S$ contains at least one code element in:
\begin{equation}
E_{F}^{c} = \{e \in E_{F}|(c, e) \in \mathsf{Evolve}\}
\label{eq:E_susp_c}
\end{equation}

Then, for each file $s \in S$, we compare the ASTs derived from $s$ before and
after the commit $c$. If the ASTs are identical for all files in $S$, we
consider the commit $c$ as a style-change commit. Note that this approach does
not guarantee 100\% recall, as it is possible for two source files to yield
different ASTs while sharing the same semantic.
However, it can safely prune the search space due to its soundness, i.e., if it
identifies a commit as a style change commit, it is guaranteed to be a style
change commit. Consequently, the search space for BIC can be further reduced to
$C_{BIC} = C_{F} \setminus C_{SC}$, in which $C_{SC}$ denotes all
identified style change commits in $C_{F}$.

\subsection{Stage 3: Scoring Commits using FL Scores and History}
\label{sec:methodology:stage3}

We are now left with the reduced BIC search space $C_{BIC}$, which only
contains commits related to the evolution of the suspicious code elements and
are also identified as non-style-change commits. The remaining task is to rank
the commits in $C_{BIC}$ in the order of their likelihood of being the BIC. As
mentioned earlier, our basic intuition is that if a commit had created, or
modified, more suspicious code elements for the observed failures, it is more
likely to be a BIC.

The suspiciousness of code elements can be quantified via an FL technique. For
example, we can apply SBFL~\cite{Wong2016} using the coverage of the test suite
$T$: note that SBFL uses only test coverage and result information, both of
which are available at the time of observing a test failure. Assuming that we
are given the suspiciousness scores, let $susp \colon E_{F} \to [0, \infty)$ be the mapping function from each suspicious code element in $E_{F}$
to its non-negative FL score.\footnote{The constraint of FL-score being
non-negative is adopted for the sake of simplicity. Note that any FL results
can be easily transformed so that the lowest score is 0.} To convert the
code-level scores to the commit level, we propose a voting-based commit scoring
model where the FL score of a code element is distributed to its relevant
commits. The model has two main components: rank-based voting power and
depth-based decay.

\begin{table}[t]
  \centering
  \caption{Example of the voting power of code elements}
  \scalebox{1.00}{
  \begin{tabular}{l|l|rrrrr}
  \toprule
  \multicolumn{2}{l|}{\textbf{Code Element}}           & $e_1$ & $e_2$ & $e_3$ & $e_4$ & $e_5$\\\midrule
  \multicolumn{2}{l|}{\textbf{Score}}            & 1.0   & 0.6   & 0.6 & 0.6 & 0.3  \\\midrule
  \multicolumn{2}{l|}{$rank_{max}$}  & 1     & 4    & 4     & 4 & 5\\
  \multicolumn{2}{l|}{$rank_{dense}$}  & 1     & 2    & 2     & 2 & 3\\\midrule
  \multirow{4}{*}{$vote$}&$\alpha=0$, $\tau=max$   & 1.00  & 0.25  & 0.25  & 0.25 & 0.20  \\
  &$\alpha=1$, $\tau=max$   & 1.00  & 0.15  & 0.15  & 0.15  & 0.06  \\
  &$\alpha=0$, $\tau=dense$ & 1.00  & 0.50  & 0.50  & 0.50  & 0.33  \\
  &$\alpha=1$, $\tau=dense$ & 1.00  & 0.30  & 0.30  & 0.30  & 0.10  \\
\bottomrule
  \end{tabular}}
  \label{tab:voting}
\end{table}

\textbf{Rank-based Voting Power:} Recent work~\cite{Sohn2019,Sohn2021,
habchi2022made} showed that, when aggregating FL scores from finer granularity
elements (e.g, statements) to a coarser level (e.g., methods), it is better to
use the \emph{relative rankings} from the original level only, rather than directly
using the scores. The actual aggregation takes the form of voting: the higher
the ranking of a code element is in the original level, the more votes it is
assigned with for the target level. Subsequently, each code element casts its
votes to the related elements in the target level. We adopt this voting-based
method to aggregate the statement level FL scores to commits. The
\emph{voting power} of each code element $e$ based on their FL rankings (and scores) as follows:

\begin{equation}
vote(e) = \frac{\alpha*susp(e) + (1 - \alpha)*1}{rank_{\tau}(e)}
\label{eq:vote}
\end{equation}

where $\alpha \in \{0, 1\}$ is a hyperparameter that decides whether to use the
suspiciousness value ($\alpha=1$) as a numerator or not ($\alpha=0$), and $\tau$
a hyperparameter that defines the tie-breaking scheme. We vary $\tau \in \{max,
dense\}$: the max tie-breaking scheme gives the lowest (worst)
rank in the tied group to all tied elements, while $dense$ gives the highest
but does not skip any ranks after ties. By design, $\tau=max$ will penalise tied
elements more severely than $\tau=dense$.
The example in Table~\ref{tab:voting} shows how the hyperparameters affect voting. Note that the relative order between FL scores is preserved in the voting power regardless of hyperparameters, i.e., $vote(e) > vote(e')$ if and only if $susp(e) > susp(e')$.

\textbf{Depth-based Decay:}
Wen et al.~\cite{Wen2016} showed that using the information about commit time
can boost the accuracy of the BIC identification. Similarly, Wu et al.~\cite{Wu2017} observed that the commit time of crash-inducing changes is closer to
the reporting time of the crashes.
Based on those findings, we hypothesise that older commits are less likely to
be responsible for the currently observed failure, because if an older commit was
a BIC, it is more likely that the resulting bug has already been found and
fixed.
To capture this intuition, we propose a depth-based decay function that
decreases the voting power of a code element as the depth of the commit in the
history of the code element increases. The historical depth of a commit $c$, with respect to a code element $e \in E_{F}^{c}$ (Eq.~\ref{eq:E_susp_c}), is defined as follows:
\begin{equation}
\begin{aligned}
depth(e,c) = & |\{c' \in C_{BIC}|\\
& (c', e) \in \mathsf{Evolve} \wedge c'.time > c.time\}|
\end{aligned}
\label{eq:depth}
\end{equation}
Note that, unlike existing work that considered the depth at the level of
commit, we consider the depth of each code element and use this to adjust the
voting power of each element.

\begin{figure}[t]
  \centerline{\includegraphics[width=\linewidth]{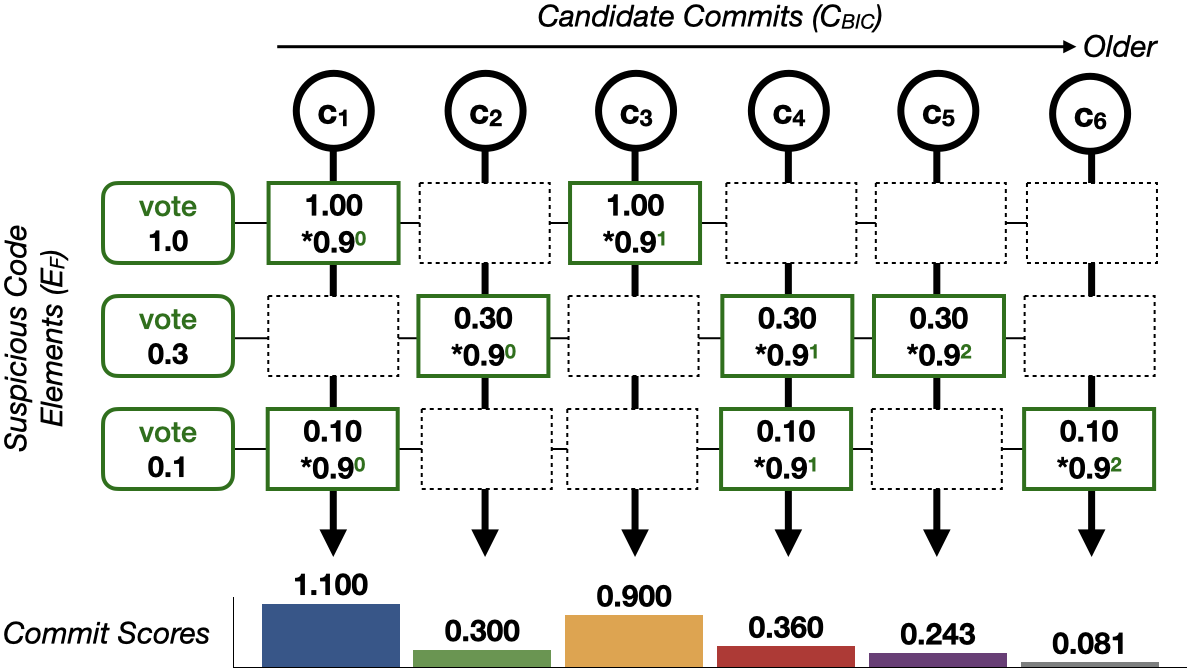}}
  \caption{Example of computing the commit scores when $\lambda = 0.1$}
  \label{fig:voting}
\end{figure}

Bringing it all together (Eq.~\ref{eq:vote} and Eq.~\ref{eq:depth}), we use the following model to assign a score to each commit $c$ in $C_{BIC}$:
\begin{equation}
  commitScore(c) = \sum_{e \in E_{F}^c}vote(e)*(1-\lambda)^{depth(e, c)}
\label{eq:commit_score}
\end{equation}
where $\lambda \in \left[0, 1\right)$ is the decay factor: when $\lambda=0$, there is no penalty for older commits. Figure~\ref{fig:voting} shows the example of calculating the score of commits when $\lambda=0.1$.

Finally, based on $commitScore$, the commit scoring function $s \colon C \to
[0, \infty)$ of \name is defined as follows:
$$s(c) =
  \begin{cases}
      commitScore(c)  & \text{if } c \in C_{BIC}\\
      0              & \text{otherwise}
  \end{cases}
$$

\section{Weighted Bisection}
\label{sec:weighted_bisection}

Bisection is a traditional way of finding the BIC by repeatedly narrowing down
the search space in half using binary search: it is implemented in popular
Version Control Systems (VSCs), e.g., \texttt{git bisect} or \texttt{svn-bisect}. A standard bisection is performed as follows: given the earliest \emph{bad}
and last \emph{good} versions, it iteratively checks whether the midpoint of
those two versions, referred to as a \emph{pivot}, contains the bug. If there
is a bug, the earliest bad point is updated to the pivot, otherwise, the last
good point is updated to the pivot. If there is a bug-revealing test case that
can automatically check the existence of a bug, the search process can be fully
automated.

However, as pointed out in previous work~\cite{Murali2021}, even though the bug
existence check can be automated, each bisect iteration may still require a
significant amount of time and computing resources, especially when the program
is large and complex, or the bug-revealing test takes a long time to execute.
Since a lengthy bisection process can block the entire debugging pipeline, we
aim to explore whether the bisection can be accelerated using the commit score
information.

\begin{figure}[t]
  \centerline{\includegraphics[width=0.95\linewidth]{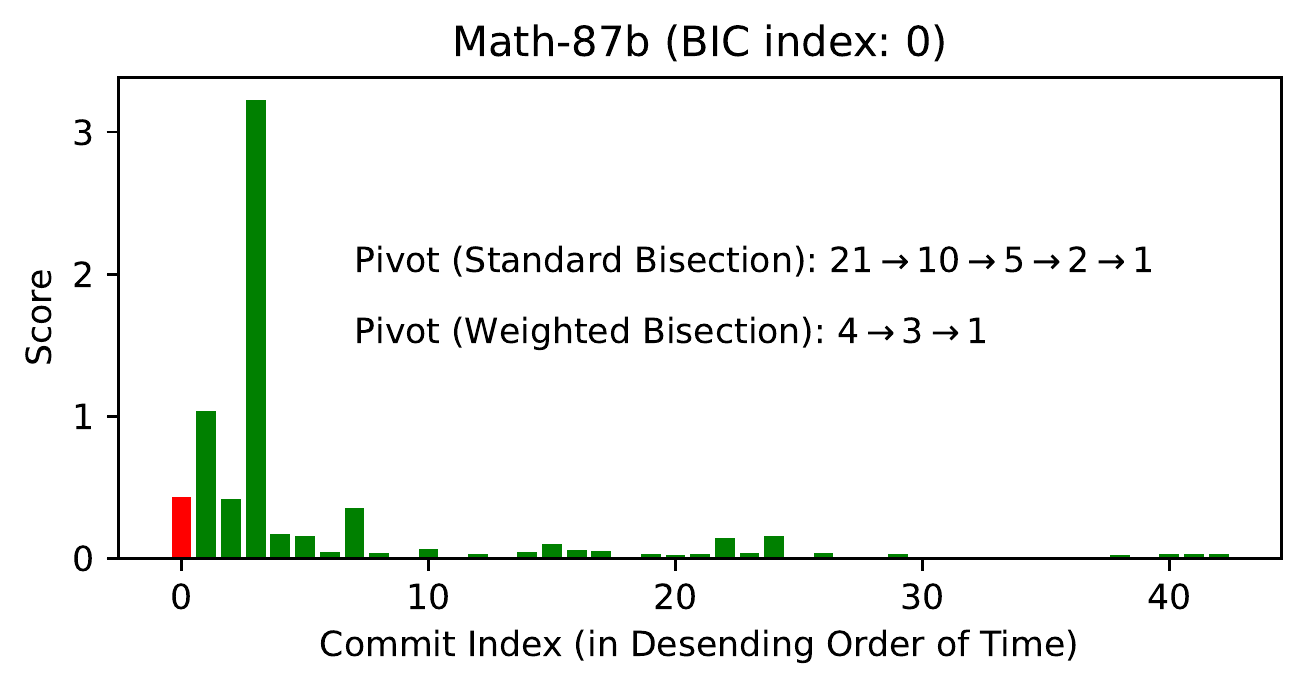}}
  \caption{Example of applying the weighted bisection to \texttt{Math-87}}
  \label{fig:search}
\end{figure}

We propose a \emph{weighted} bisection algorithm, where the search pivot is
set to a commit that will halve \emph{the amount of remaining commit scores}
instead of \emph{the number of remaining commits}, in order to reduce the
number of highly suspicious commits more quickly. For example, let us consider
the example in Fig.~\ref{fig:search} that shows the score distribution of the
commits in the reduced BIC search space of \texttt{Math-87} in Defects4J. For
\texttt{Math-87}, the score distribution is biased towards a small number of
recent commits including the real BIC (marked in red) with the third highest score. In this
case, simply using the midpoint as a search pivot might not be a good choice
because all highly suspicious commits still remain together on one side of the
split search space: as a result, the standard bisection requires five iterations to finish.
Alternatively, if we pivot at the commit that halves the amount of
remaining scores, the bisection reaches the actual BIC more quickly, completing
the search in three iterations. Note that this algorithm is a \emph{generalised}
version of the standard bisection: the standard bisection is a special case of
weighted bisection with all commits sharing the same non-zero score.

Algorithm~\ref{algo:search} presents the weighted bisection algorithm.
It takes the set of commits $C$, and the commit score function $s \in C
\rightarrow [0, \infty)$, as input, and returns the BIC.
First, it removes all commits with a score of zero (Line 1), and sorts the
remaining commits in the descending order of their commit time (Line 2).
Assuming that there is at least one BIC in the sorted sequence $C'$, the
earliest bad index $bad$ is set to 0, the index of the most recent commit (Line
3). Since all commits in $C'$ are BIC candidates, we set the last good index
$good$ to the index just after the oldest commit (Line 4). Then, a new $pivot$
index is iteratively selected from the range $[bad+1, good-1]$, until there is
no remaining commit between $bad$ and $good$ (Line 5). As mentioned earlier, we
select a pivot that minimises the difference between the left (not including
pivot) and the right (including pivot) sum of the scores (Line 6). Once a new
pivot is selected, the commit $C'[pivot]$ is inspected for the bug, either using the bug-revealing tests or manually (Line
7). If a bug is detected, the $bad$ index is updated to $pivot$ (Line 8),
otherwise, the $good$ index is updated to $pivot$ (Line 10). Finally, it
returns the identified BIC at the $bad$ index when the loop terminates (Line
11).

\begin{algorithm}[t]
  \small
  \SetCommentSty{mycommfont}
  \SetKwInput{KwPrecondition}{Precondition}
  \SetKwInput{KwPostcondition}{Postcondition}
  \KwIn{Set of commits $C$}
  \KwIn{Commit score (weight) function $s: C \rightarrow [0, \infty)$}
  \KwOut{Bug inducing commit $c \in C$}
  \tcp{Remove irrelevant commits}
  $C \leftarrow \{c \in C|s(c) > 0\}$\\
  $C' \leftarrow C\mathsf{.orderByTimeDesc()}$\\
  \tcp{$C'[i]$ is newer than $C'[i+1]$}
  $bad \leftarrow 0$\\
  $good \leftarrow |C'|$\\
  \While{$good > bad + 1$}{
    \tcp{$\mathsf{S}(a, b) = \sum_{i=a}^{b}s(C'[i])$}
    $pivot \leftarrow \text{argmin}_{i=bad+1}^{good-1}{|\mathsf{S}(bad, i-1) - \mathsf{S}(i, good-1)|}$\\
    \If{$C'[pivot]\mathsf{.ContainsBug()}$}{
      $bad \leftarrow pivot$\\
    }\Else{
      $good \leftarrow pivot$\\
    }
  }

  \tcp{assert $bad = good - 1$}

  \Return{$C'[bad]$}
  \caption{Weighted Bisection Algorithm}
  \label{algo:search}
\end{algorithm}

\section{Evaluation Setup}
\label{sec:eval_setup}

This section presents our research questions and describes the experimental
setup.

\subsection{Research Questions}
We ask the following research questions in this paper:
\begin{itemize}
\item\textbf{RQ1}: How accurately does \name rank the BIC?
\item\textbf{RQ2}: How efficient is the weighted bisection compared to the standard bisection?
\item\textbf{RQ3}: What is the impact of FL accuracy to the performance of \name?
\end{itemize}

\subsection{Dataset of Bug Inducing Commits}
\label{sec:dataset}
We choose Defects4J v2.0.0~\cite{Just2014}, a collection of 835 real-world bugs
in Java open-source programs, as the source of our experimental subjects.
While Defects4J provides test suites containing the bug-revealing tests
for every bug, as well as the entire commit history for each buggy version, it
lacks the BIC information for each bug. We, therefore, start with a
readily-available BIC dataset for 91 Defects4J bugs\footnote{https://github.com/justinwm/InduceBenchmark} constructed by Wen et al.~\cite{Wen2019}. This
dataset was created by running the bug-revealing test cases on the past
versions and finding the earliest buggy version that makes the tests fail.
However, in our experiment, we are forced to exclude 24 out of 91 data points.
Since \name is implemented using Git, it cannot trace the commit history of
nine bugs from the \texttt{JFreeChart} project which uses SVN as its version
control system. Further, we exclude 14 data points that are shown to be
inaccurate by previous work~\cite{An2021}. Lastly, \texttt{Time-23} is also
discarded, because we found that the identified commit in the dataset does not
contain any change to code, but only to the license comments. The detailed
reasons can be found in our repository.

We augment the remaining 67 BIC data points from Wen et al.~\cite{Wen2019} by
inspecting some of the remaining bugs in Defects4J. Two of the authors
manually and independently identified the BIC for each bug, consulting the bug
reports, failure symptoms, and developer patches. To reduce the manual
inspection cost, we only targeted the bugs for which the cardinality of the
reduced BIC search space, $C_{BIC}$, is less than or equal to ten. Two authors
reached on consensus for 70 data points, which have been added to the dataset.
In summary, a total of 130 data points (67 from Wen et al.~\cite{Wen2019} + 70
manually curated - 7 overlapped) are used for the evaluation of \name. The
combined BIC dataset and the provenance of each data point are available in our
repository for further scrutiny.

\subsection{Implementation Details}

\begin{table}[t]
  \centering
  \caption{Example of Relevant Test Selection (\texttt{Time-15})}
  \scalebox{0.95}{
  \begin{tabular}{l}
  \toprule
  \textbf{Failing Test ($T_F$)}\\\midrule
  org.joda.time.field.TestFieldUtils::testSafeMultiplyLongInt\\\midrule
  \textbf{Classes Covered by the Failing Test}\\\midrule
  org.joda.time.field.\textbf{FieldUtils}\\
  org.joda.time.\textbf{IllegalFieldValueException}\\\midrule
  \textbf{Relevant Tests ($T \setminus T_F$)}\\\midrule
  org.joda.time.Test\textbf{IllegalFieldValueException}::testGJCutover\\
  org.joda.time.Test\textbf{IllegalFieldValueException}::testJulianYearZero\\
  org.joda.time.Test\textbf{IllegalFieldValueException}::testOtherConstructors\\
  org.joda.time.Test\textbf{IllegalFieldValueException}::testReadablePartialValidate\\
  org.joda.time.Test\textbf{IllegalFieldValueException}::testSetText\\
  org.joda.time.Test\textbf{IllegalFieldValueException}::testSkipDateTimeField\\
  org.joda.time.Test\textbf{IllegalFieldValueException}::testVerifyValueBounds\\
  org.joda.time.Test\textbf{IllegalFieldValueException}::testZoneTransition\\
  org.joda.time.field.Test\textbf{FieldUtils}::testSafeAddInt\\
  org.joda.time.field.Test\textbf{FieldUtils}::testSafeAddLong\\
  org.joda.time.field.Test\textbf{FieldUtils}::testSafeMultiplyLongLong\\
  org.joda.time.field.Test\textbf{FieldUtils}::testSafeSubtractLong\\
  \bottomrule
  \end{tabular}}
  \label{tab:relevant}
\end{table}

We apply \name at the \emph{statement}-level granularity, i.e., $E$ is a set
of statements composing the target buggy program. The initial BIC search space,
$C$, is set to all commits from the very first commit up to the commit
correspond to the buggy version, i.e., \texttt{revision.id.buggy} in
Defects4J. Among the given test suite in Defects4J, we only use the bug-revealing (i.e.,
failing) test cases as well as their relevant test cases as $T$. A test case is
relevant if and only if its full name contains the name of at least one class
executed by the failing test cases. Table~\ref{tab:relevant} shows the example
of the relevant test selection.

\subsubsection{Construction of the $\mathsf{Cover}$ relation}
To construct the $\mathsf{Cover}$ relation between $T$ and $E$, we measure the
statement-level coverage of each test case in $T$ using \texttt{Cobertura v2.0.3} which
is included in Defects4J.

\subsubsection{Construction of the $\mathsf{Evolve}$ relation}
To construct the $\mathsf{Evolve}$ relation between $C$ and $E$, we need to
retrieve the commit history of each code element: we use the \texttt{git log} command\footnote{\texttt{git log -C -M -L<start\_line>,<end\_line>:<file>}. The options \texttt{-C} and \texttt{-M} detect file rename/copy/move between versions.} following our previous work~\cite{An2021}. We also attempted using CodeShovel~\cite{Grund2021}, a state-of-the-art method history retrieval tool, instead of \texttt{git log}, but found that the tool sometimes produces incorrect histories. Since it is infeasible to manually validate all commits retrieved by CodeShovel, we only report the results with \texttt{git log} in this paper. However, we include the commit history retrieved by CodeShovel (with the incorrect outputs) in our artefact for further validation and comparison.

Please note that for each statement, we retrieve the commit history of its
enclosing method and create the $\mathsf{Evolve}$ relations between the
statements and the retrieved commits to ensure high recall for commit
histories. This is also to deal with omission bugs~\cite{Zhang2007}: if a bug is caused by
omission of some statements, we cannot trace the log of the missing statements
because they literally do not exist in the current version. In that case,
tracing the log of the neighbouring statements (in the enclosing method) will
enable to find the inducing commit, as the method that encloses the omission
bug should have been covered by the failing tests~\cite{An2021}.


\subsubsection{Detection of Style-Change Commits}
For Stage 2, we use OpenRewrite v7.21.0\footnote{https://github.com/openrewrite/rewrite} to ensure the same coding standard between the two versions of files. More
specifically, we use the \texttt{Cleanup} recipe\footnote{https://docs.openrewrite.org/reference/recipes/java/cleanup} that fixes any errors that violate CheckStyle rules.\footnote{https://checkstyle.sourceforge.io/}
This ensures that trivial differences between two versions that do not lead to
semantic differences are ignored: a good example is a commit in \texttt{Lang},
which is shown in Fig.~\ref{fig:Lang-46b-5814f50}. To compare ASTs, we use the
isomorphism test of GumTree v3.0.0~\cite{Falleri2014} that has time
complexity of $O(1)$.

\subsubsection{Fault Localisation}

To obtain the FL score of each statement, we use a widely-used SBFL formula, Ochiai~\cite{Abreu2006}, which can be expressed in our context as follows:
$$
Ochiai(e) = \frac{|\{t \in T_F|(t, e) \in \mathsf{Cover}\}|}{\sqrt{|T_F|*|\{t \in T|(t, e) \in \mathsf{Cover}\}|}}
$$
By definition, $Ochiai(e) > 0$ if and only if $e \in E_{F}$ (Eq.~\ref{eq:E_susp}).

\subsection{Evaluation Metrics}
If the scoring model works well, BICs will have higher scores and ranks than
other commits. Therefore, we use the following widely-adopted ranking-based
evaluation metrics. When there are tied elements, the max-tiebreaker is used.
\begin{itemize}
 \item Accuracy@n: The number of subjects where the ranking of the BIC is within the top $n$ positions (\emph{higher is better})
 \item Mean Reciprocal Rank (MRR)~\cite{Craswell2009}: The average reciprocal rank of the BIC (\emph{higher is better})
\end{itemize}

\subsection{Baselines}
We compare \name to the following baselines.
\subsubsection{Other Voting Schemes}
\begin{itemize}
\item Equal: All lines that are covered by failing test cases are assigned the same weight, i.e., \inlineequation[eq:vote_equal]{vote(e) = 1}.
\item Only Score: The voting power of a method is simply defined as its FL score without considering the ranking, i.e., \inlineequation[eq:vote_score]{vote(e) = susp(e)}.
\end{itemize}

\subsubsection{Other Scoring/Ranking Techniques}
\begin{itemize}
\item Random: This strategy randomly shuffles the commits in the search space. The random strategy is a meaningful baseline for ranking-based evaluation because the ranking result can be overestimated when the size of the search space is small. When there are $n$ commits in the search space, the expected rank of the BIC is $\frac{1 + n}{2}$.
\item Max (Eq.~\ref{eq:score_max}): Instead of Eq.~\ref{eq:commit_score}, the score of a commit is defined as the highest suspiciousness score of code elements that are modified by the commit:
\begin{equation}
  commitScore(c) = \max_{e \in E_{F}^c}susp(e)
\label{eq:score_max}
\end{equation}
Similarly, in Orca~\cite{Bhagwan2018}, the file-level scores are converted into the commit level using max-aggregation. Many FL techniques have used this scheme when the granularity of the code elements in the coverage matrix and the target FL granularity are different~\cite{Sohn2017, Lou2020}.
\item FBL-BERT~\cite{Ciborowska2022}: FBL-BERT is a recently proposed changeset
localisation technique based on a pre-trained BERT model called
BERTOverflow~\cite{tabassum2020code}. Given a bug report, it retrieves the
relevant changesets using their scores obtained by the BERT-based model. We
fine-tune the model using the training dataset from the \texttt{JDT} project,
which is the largest training dataset provided by their repository\footnote{We
confirm that the model fine-tuned with \texttt{JDT} performs better than that
fine-tuned with \texttt{ZXing}, which has the smallest training dataset.}: this
is because no such training data is available for our target projects. We use
the ARC changeset encoding strategy, which performed the best for
changeset-level retrieval in the original study~\cite{Ciborowska2022}. As
Defects4J contains the link to the original bug report for every bug, we use
the contents of the original bug report as an input query.
\item Bug2Commit~\cite{Murali2021}: Bug2Commit is a state-of-the-art IR-based
BIC identification method for large-scale systems: it exploits multiple
features of commits and bug reports. When implementing Bug2Commit, we use the
Vector Space Model (VSM) because the word-embedding model requires an
additional training dataset of bug reports and commits. As in the original
paper, we use BM25~\cite{robertson1995okapi} as a vectoriser and use the Ronin
tokeniser, the most advanced splitter in \texttt{Spiral}~\cite{spiral2018}\footnote{https://github.com/casics/spiral}. We use two features of commit: the commit
message and the modified file names. From bug reports, we use three features:
the exception message and stack traces from failing test cases, the title of
bug report, and the content of bug report.
\end{itemize}

\section{Results}
\label{sec:results}
This section presents the results of our experiments.

\subsection{\textbf{RQ1}: Ranking Performance of \name}

\begin{figure}[t]
  \centerline{\includegraphics[width=0.95\linewidth]{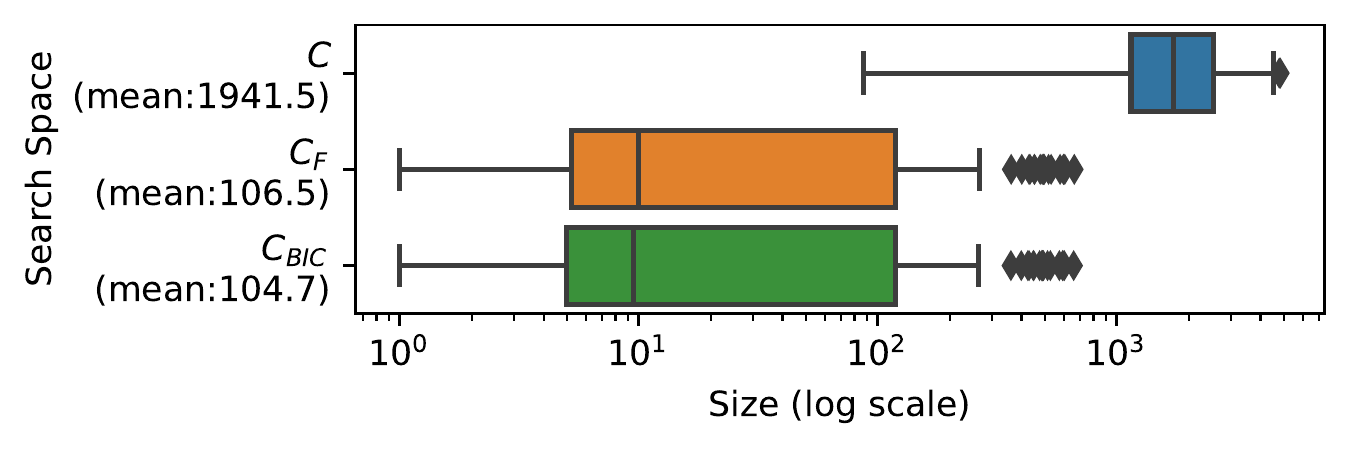}}
  \caption{Distributions of the sizes of search space}
  \label{fig:RQ1-SS}
\end{figure}

\begin{table}[t]
  \centering
  \caption{The distribution of the size of the reduced search space}
  \scalebox{0.90}{
  \begin{tabular}{l|rrrrrrr}
    \toprule
    $|C_{BIC}|$ &   $\leq 1$ &   $\leq 2$ &   $\leq 3$ &   $\leq 5$ &  $\leq 10$ &  $\leq 20$ &  $\leq 30$\\\midrule
    \# Subjects &          3 &          8 &         19 &         41 &         71 &         76 &         76\\\midrule
    $|C_{BIC}|$ &  $\leq 50$ & $\leq 100$ & $\leq 200$ & $\leq 300$ & $\leq 500$ & $\leq 600$ & $\leq 700$\\\midrule
    \# Subjects &         83 &         94 &        106 &        112 &        123 &        128 &        130\\\bottomrule
  \end{tabular}}
  \label{tab:RQ1-SS}
\end{table}

Let us first check how much search space reduction is achieved by Stages 1 and 2
of \name. Figure~\ref{fig:RQ1-SS} shows the distribution of the sizes of $C$
(original), $C_{F}$ (after Stage 1), and $C_{BIC}$ (after Stage 1+2),
respectively, over all subjects. The results show that Stage 1 significantly
reduces the size of the search space. On average, the size of search space can
be reduced to 7\% of its original size using only the coverage of the failing
tests.
The search space reduction by Stage 2 is relatively marginal compared to Stage 1: the
style change commits are detected in only 75 out of 130 subjects (58\%). To
make it easier for readers to grasp the ranking-based evaluation results
that follow, we report the size distribution of the final reduced search
space $C_{BIC}$ in Table~\ref{tab:RQ1-SS}.

\begin{figure}[t]
  \centerline{\includegraphics[width=\linewidth]{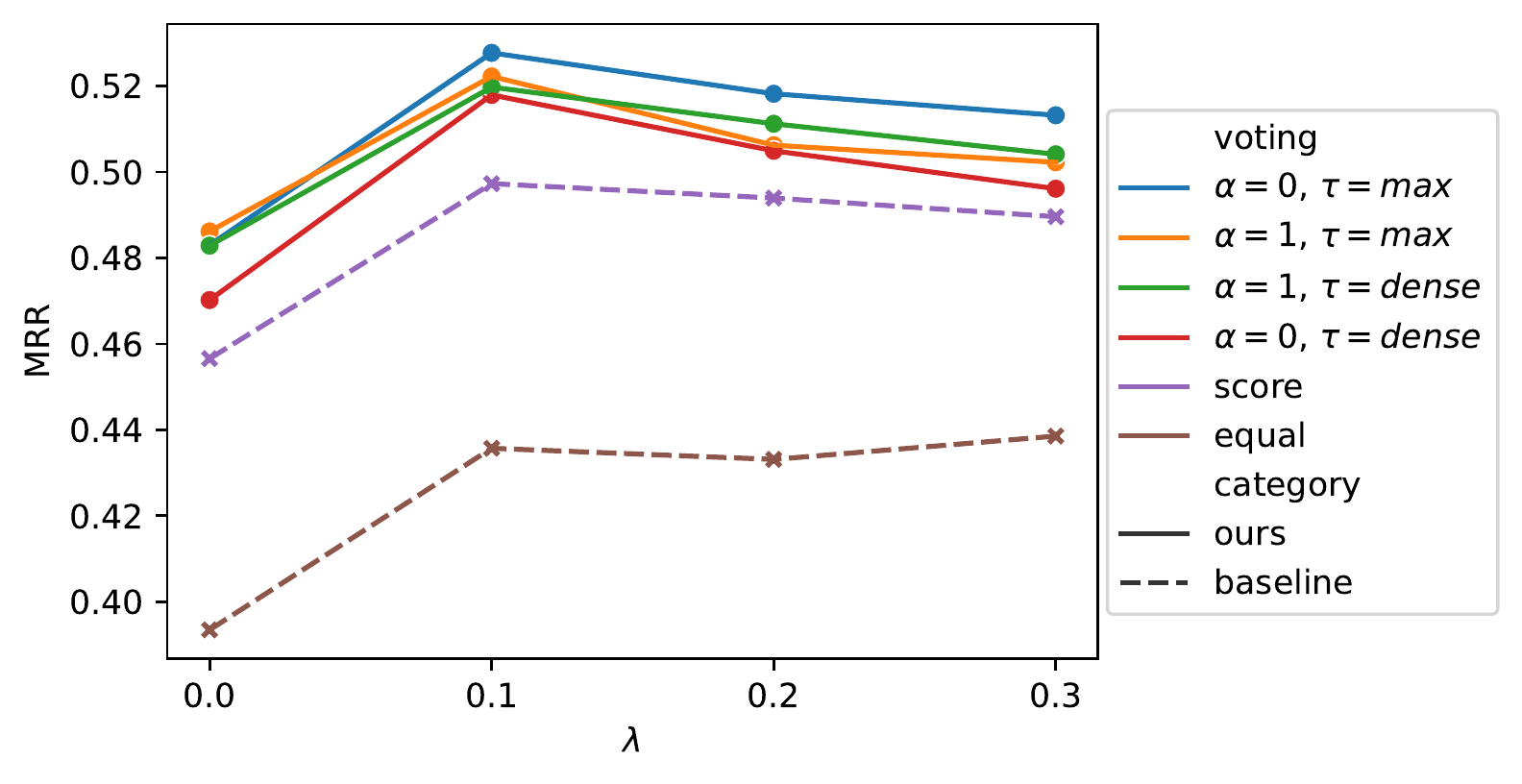}}
  \caption{MRR for each hyperparameter configuration of \name}
  \label{fig:RQ1-MRR}
\end{figure}

\begin{table*}[t]
  \centering
  \caption{Comparison of the performance of \name (with $\alpha=0, \tau=\text{max}, \lambda=0.1$) to other commit ranking techniques applied to two commit search space $C_{BIC}$ (after Stages 1 and 2 of \name) and $C$, respectively. The evaluation is performed on two sets of subjects: the dataset from Wen et al. (67 subjects) and our manually curated set of 63 subjects. The performance was measured using MRR (Mean Reciprocal Rank) and accuracy@k, where k is 1, 2, 3, 5, and 10.}
  \scalebox{0.87}{
    \begin{tabular}{l|r|rrrrr|r|rrrrr|r|rrrrr}
      \toprule
      Subjects & \multicolumn{6}{c|}{\textbf{All (\# subjects = 130)}} & \multicolumn{6}{c|}{From Wen et al.~\cite{Wen2019} (\# subjects = 67)} & \multicolumn{6}{c}{Manually Curated (\# subjects = 63)}\\\midrule
      \multirow{2}{*}{Metric} & \multirow{2}{*}{MRR} & \multicolumn{5}{c|}{Accuracy} & \multirow{2}{*}{MRR} & \multicolumn{5}{c|}{Accuracy} & \multirow{2}{*}{MRR} & \multicolumn{5}{c}{Accuracy} \\\cmidrule{3-7}\cmidrule{9-13}\cmidrule{15-19}
          &       &     @1 &     @2 &     @3 &     @5 &     @10 &       &     @1 &     @2 &     @3 &     @5 &     @10 &       &     @1 &     @2 &     @3 &     @5 &     @10 \\\midrule
      \multirow{2}{*}{\name} & \multirow{2}{*}{\textbf{0.528}} &     \textbf{47} &     \textbf{66} &     \textbf{85} &     \textbf{98} &     \textbf{110} & \multirow{2}{*}{0.324} &     \multirow{2}{*}{ 9} &     \multirow{2}{*}{19} &     \multirow{2}{*}{29} &     \multirow{2}{*}{38} &      \multirow{2}{*}{47} & \multirow{2}{*}{0.745} &     \multirow{2}{*}{38} &     \multirow{2}{*}{47} &     \multirow{2}{*}{56} &     \multirow{2}{*}{60} &      \multirow{2}{*}{63} \\
                       &               & {\scriptsize (36\%)} &	{\scriptsize (51\%)} &	{\scriptsize (65\%)} &	{\scriptsize (75\%)} &	{\scriptsize (85\%)} & &  &  &  &  &  & &  &  &  &  &  \\
                  \midrule
    \multicolumn{19}{l}{\textbf{Other Techniques (on $C_{BIC}$)}} \\\midrule
            Bug2Commit & 0.380 & 27 & 42 & 64 & 85 & 96& 0.235 &      7 &     13 &     19 &     26 &      33  & 0.534 &     20 &     29 &     45 &     59 &      63\\
            FBL-BERT & 0.338 & 27 & 40 & 47 & 69 & 90& 0.158 &      5 &      9 &     11 &     14 &      27  & 0.529 &     22 &     31 &     36 &     55 &      63\\

            Random Baseline& 0.218 &      3 &     19 &     41 &     65 &      75 & 0.065 &      0 &      2 &      4 &      6 &      12  & 0.381 &      3 &     17 &     37 &     59 &      63\\
            Theoretical Lower Bound & 0.145 &      3 &      8 &     19 &     41 &      71 & 0.039 &      0 &      1 &      2 &      4 &       8  & 0.258 &      3 &      7 &     17 &     37 &      63\\\midrule
        \multicolumn{19}{l}{\textbf{Other Techniques (on $C$)}} \\\midrule
          Bug2Commit & 0.155 & 11 & 18 & 22 & 25 & 39 & 0.123 &      4 &      7 &      9 &     11 &      16 & 0.189 &      7 &     11 &     13 &     14 &      23\\
          FBL-BERT & 0.037 & 1 & 3 & 5 & 7 & 10 & 0.037 &      1 &      2 &      2 &      3 &       4 & 0.036 &      0 &      1 &      3 &      4 &       6\\
          Random Baseline & 0.002 &      0 &      0 &      0 &      0 &       0 & 0.002 &      0 &      0 &      0 &      0 &       0 & 0.002 &      0 &      0 &      0 &      0 &       0\\
          Theoretical Lower Bound & 0.001 &      0 &      0 &      0 &      0 &       0 & 0.001 &      0 &      0 &      0 &      0 &       0 & 0.001 &      0 &      0 &      0 &      0 &       0\\\midrule
          \multicolumn{19}{l}{\textbf{Ablation Study for \name}} \\\midrule
          Skip Stage 2 & 0.490 &     39 &     64 &     82 &     97 &     110  & 0.317 &      9 &     18 &     28 &     37 &      47 & 0.675 &     30 &     46 &     54 &    60 &      63 \\
          Use Equal Vote (No FL) & 0.436 &     39 &     56 &     67 &     79 &      88  & 0.193 &      7 &      9 &     12 &     19 &      25 & 0.694 &     32 &     47 &     55 &    60 &      63 \\
          Max Aggr. (Eq.~\ref{eq:score_max}) & 0.317 &     17 &     36 &     50 &     73 &      97 & 0.142 &      0 &      5 &      9 &     18 &      34  & 0.503 &     17 &     31 &     41 &     55 &      63\\
      \bottomrule
    \end{tabular}}
  \label{tab:RQ1-ranking}
\end{table*}

We now turn to the ranking performance of \name: Fig.~\ref{fig:RQ1-MRR}
presents the MRR metric achieved by \name with each hyperparameter setting.
The MRRs from our ranking-based voting power (Eq.~\ref{eq:vote}) are plotted in
solid lines, while those from the baseline voting schemes,
(Eq.~\ref{eq:vote_equal}, and Eq.~\ref{eq:vote_score}), are plotted in dashed
lines. We observe that any hyperparameter setting of \name can outperform the
baseline voting methods, which demonstrates the effectiveness of using the FL
rank of code elements in allocating voting power. Regarding the depth-based
decay of voting power, we observe that decay weights $\lambda \in \{0.1, 0.2,
0.3\}$ perform better than $\lambda = 0.0$ (i.e., no decay). In particular,
setting $\lambda$ to $0.1$ consistently outperforms other combinations of $\tau$
and $\alpha$.

Table~\ref{tab:RQ1-ranking} shows the comparison between the performance of \name
with its best hyperparameter setting ($\alpha=0$, $\tau=\text{max}$, $\lambda=0.1$) and other baseline ranking techniques, Bug2Commit and FBL-BERT, in ranking commits in $C_{BIC}$ for all subjects. We provide the breakdown of the results based on the source of the datasets, Wen et al.~\cite{Wen2019} and our manual curation (see Section~\ref{sec:dataset}), because they have different size distributions of the reduced search space $C_{BIC}$. Our manually created dataset contains only the subjects with $|C_{BIC}| \leq 10$, so that the worst rank of the BIC in $C_{BIC}$ is still within the top 10.
Furthermore, in addition to the performance of Bug2Commit and FBL-BERT, we also provide a random baseline, which involves randomly shuffling the commits in $C_{BIC}$ and ranking them, and a theoretical lower bound, which assigns the worst possible rank to the actual BIC, to assist readers in comprehending the results for each of the datasets.

The results show that \name surpasses all other baseline techniques in terms of evaluation metrics, with a 39\% and 56\% higher MRR compared to Bug2Commit and FBL-BERT, respectively, and also achieves the highest accuracy for all studied $n$ values. As the size of the search space, $|C_{BIC}|$, increases\footnote{The size of the search space can be estimated by the theoretical lower bound of the performance.}, the BIC identification problem becomes more challenging, which is reflected in the lower performance of all techniques in the dataset from Wen et al.~\cite{Wen2019} compared to the manually curated dataset. Nonetheless, \name outperforms other techniques regardless of the size of $C_{BIC}$.

To study how much contribution the search space reduction (Stage 1 and 2 of \name) makes, we also present the performance of Bug2Commit and FBL-BERT when applied to the entire commit search space, $C$. The results show a significant decrease in the ranking performance of both techniques when the search space is not reduced before ranking.

Furthermore, the ablation experiments (\textbf{Ablation Study for \name} in Table~\ref{tab:RQ1-ranking}) show that filtering out the style change
commits can increase the MRR of \name by 8\%, showing that Stage 2 is making a
significant contribution to the ranking despite the marginal contribution to
the search space reduction. However, also note that \name still outperforms all
other baseline techniques even without Stage 2, which shows the effectiveness of
our voting-based scoring model with SBFL. Interestingly, the equal voting model (Eq.~\ref{eq:vote_equal}) with $\lambda=0.1$, which does not leverage any FL
result except the failure coverage, yet performs better than all ranking
baselines in terms of MRR. This shows that even without using the FL
techniques, simply giving equal voting power to every code element covered by
the failing test cases can rank BICs more effectively than the baselines.
On the other hand, given the same FL results with \name, the max-aggregation method achieves only 60\% of MRR compared to \name, demonstrating the importance of the voting-based aggregation.

\noindent\textbf{Answer to RQ1}: \name can rank the actual BICs in the top 1 and 5 commits for 36\% and 75\% of studied bugs. It is significantly more effective than the random baseline and the state-of-the-art IR-based BIC ranking method.

\subsection{\textbf{RQ2}: Standard Bisection vs. Weighted Bisection}
\begin{figure}[t]
  \centerline{\includegraphics[width=\linewidth]{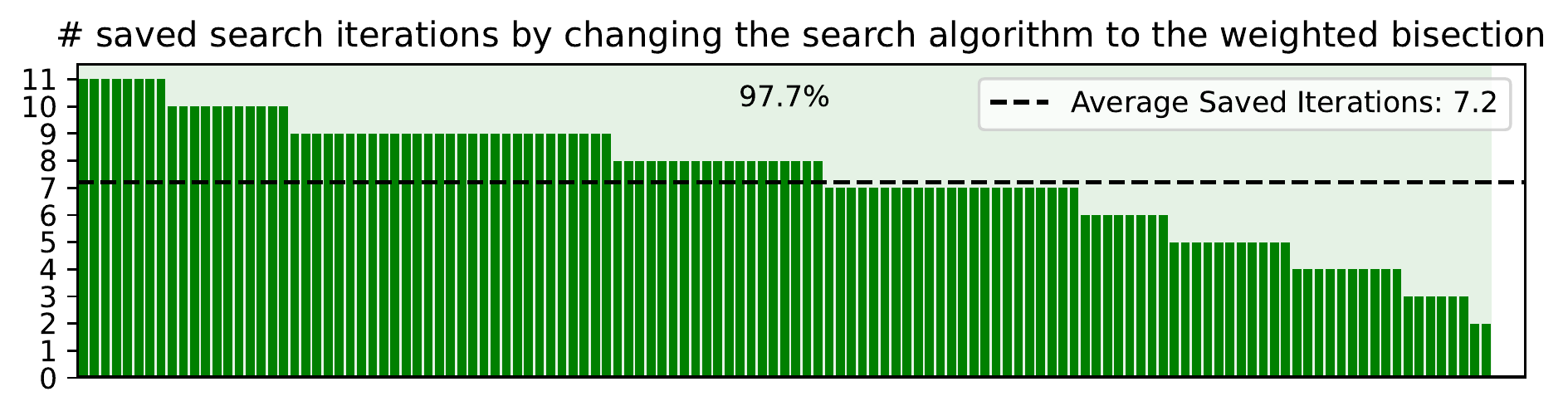}}
  \caption{The number of saved search iterations required until finding the BIC using the weighted bisection compared to the standard bisection on the \textbf{entire} commit history, $C$}
  \label{fig:RQ2-all}
\end{figure}

We simulate the standard and weighted bisection algorithms on all subjects,
assuming that the bug-revealing tests can perfectly reveal the existence of
bugs. Fig.~\ref{fig:RQ2-all} contains a sorted bar chart that shows the number
of saved search iterations, for all subjects, until finding the BIC using our
weighted bisection algorithm compared to the standard bisection on the entire
commit history. The results show that using the weighted bisection with
\name-generated scores\footnote{\name with $\alpha=
0$, $\tau=max$, $\lambda=0.1$} can reduce the search cost for about 98\% of subjects compared
to the standard bisection while saving up to 11 search iterations. On average, the number of iterations is reduced by 67\%. There is no
case where the weighted bisection degrades the performance.

\begin{figure}[t]
  \centerline{\includegraphics[width=\linewidth]{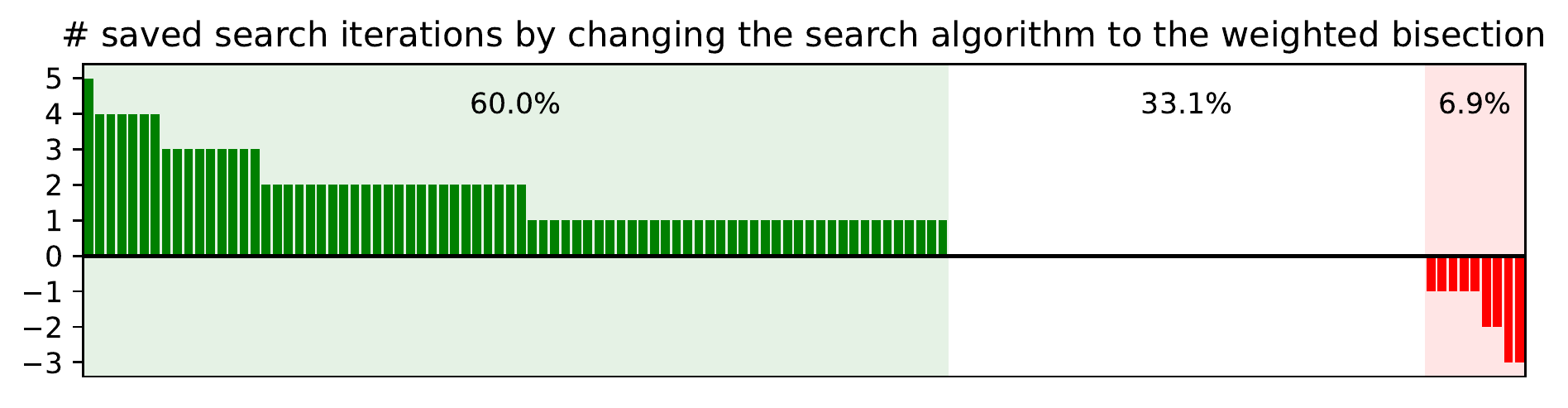}}
  \caption{The number of saved search iterations required until finding the BIC using the weighted bisection compared to the standard bisection on the \textbf{reduced} commit history, $C_{BIC}$}
  \label{fig:RQ2-reduced}
\end{figure}

For a more conservative comparison, we also compare the weighted bisection to
the standard bisection when both are applied to the reduced search space,
$C_{BIC}$.
Fig.~\ref{fig:RQ2-reduced} shows that the weighted bisection can reduce the
number of required search iterations for 78 out of 130 subjects (60.0\%), while
the number of iterations is increased in only nine out of 130 subjects (6.9\%).
In the remaining 33.1\% of cases, the number of iterations is the same as the
standard bisection.
To ensure that the median of the number of saved iterations is positive (which
would indicate that there is a \emph{performance improvement}), we perform
the one-sided Wilcoxon signed rank test~\cite{Wilcoxon1992}, whose null hypothesis is that the
median of is negative (\emph{performance degradation}). The p-value is
$1.51*10^{-11}$, allowing us to reject the null hypothesis in favour of the
alternative that \emph{the median of the number of saved iterations is
greater than zero}.

We also investigate why the weighted bisection worsens the search efficiency
for those nine subjects (6.9\%) and report that the BIC is not ranked well in
the cases, i.e. either not in the top 10 or even top 50\%. The BIC rank (in
percentage) and the number of saved iterations are negatively correlated with
each other with a Pearson correlation coefficient of -0.58.


\noindent\textbf{Answer to RQ2}: Weighted bisection combined with \name can
save the search cost in 98\% of studied bugs compared to the standard bisection
applied to the entire commit history, saving 7.2 inspections on average.
When the bisection is performed only with the reduced candidates, weighted
bisection saves the number of search iterations in 60\% of cases while
increasing it in only 7\% of cases with lower BIC ranks.

\subsection{\textbf{RQ3}: Impact of FL Accuracy on \name}

To see how the accuracy of FL affects the performance of \name on each
individual subject, we provide less accurate FL results to \name and observe
how it affects the ranking performance. We intentionally weaken the test suite
by removing some of the relevant passing test cases, as it is known that the
accuracy of SBFL is highly dependent on the quality of the used test
suite~\cite{Perez2017}. By doing so, we limit the test suite to only the test
cases that are contained in the failing test classes. For example, in the case
of Table~\ref{tab:relevant}, the relevant test cases are limited to the last
four test cases that are in the \texttt{TestFieldUtils} class containing the
failing test case.

\begin{figure}[t]
  \centerline{\includegraphics[width=0.95\linewidth]{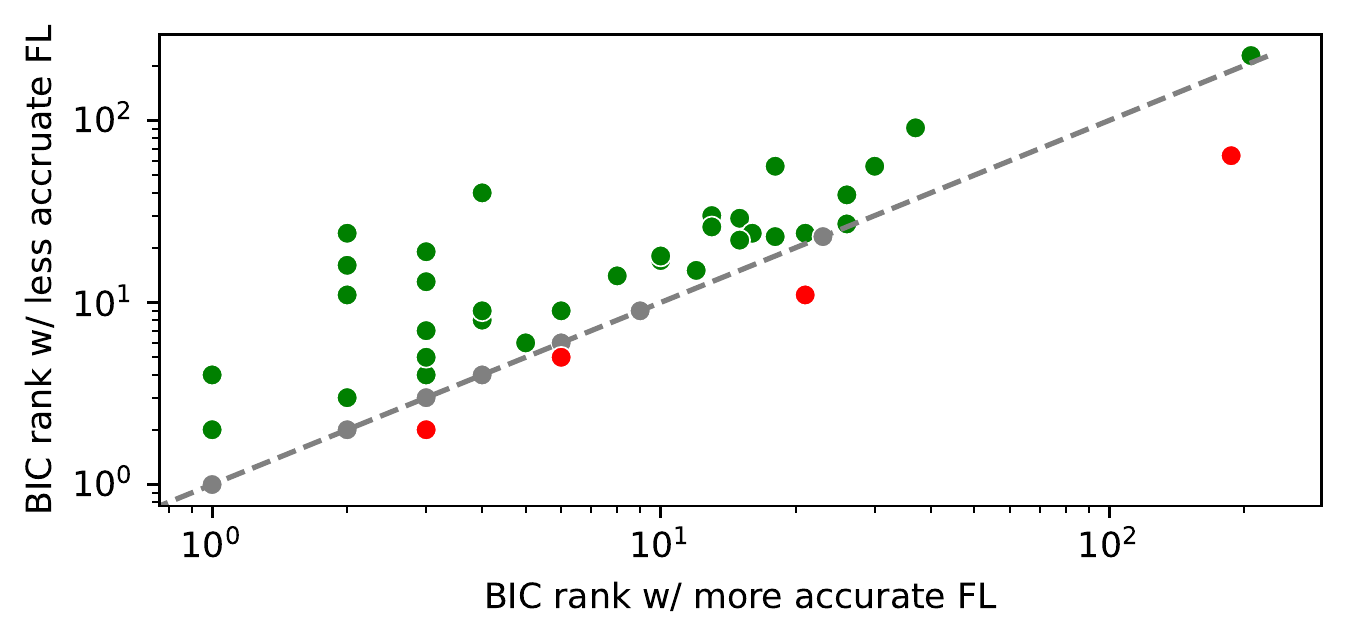}}
  \caption{BIC ranks of \name with the more and less accurate FL results}
  \label{fig:RQ3}
\end{figure}

Among the 99 out of 130 subjects whose sets of relevant test cases are reduced,
we observe that, in 57 subjects, the FL accuracy (in terms of the highest rank
of buggy methods) is decreased as a result. For those 57 subjects, we see
whether the performance of \name is affected by the accuracy of FL. In
Figure~\ref{fig:RQ3}, the $x$- and $y$-axis represent the BIC ranks produced by
\name with the more (original) and less accurate FL results, respectively.
Green markers (above the dashed line) represent the cases where the better FL
yields the better BIC rank, while red markers indicate the opposite. The overall
tendency is that higher FL accuracy leads to a better ranking performance of
\name, as shown by the fact that the number of green dots above the dotted line
is much higher than the number of red dots below the line.\footnote{\name with the worse FL results still outperforms all baselines in RQ1.}
The one-sided Wilcoxon signed rank test for the paired ranking samples also
results in the p-value of $2.56 * 10^{-6}$ showing that the median rank
difference is greater than zero when the FL accuracy increases.

\noindent\textbf{Answer to RQ3}: \name performs better when the FL results used
as its input become better. Consequently, we expect that \name can benefit from
more precise and sophisticated FL techniques in the future.

\section{Application to Industry Software}
\label{sec:industry}

\product is a large-scale
commercial software that consists of more than 10M lines of C++ and C. In the
CI system of \product, multiple commits that have individually passed the
pre-submit testing are merged into the delivery branch and tested together using a more extensive test suite on a daily basis. Considering
the set of multiple commits as a single batch, this is a type of \emph{Batch
Testing}~\cite{Najafi2019}. While batch testing reduces the overall test
execution cost for \product, it also has some practical drawbacks: when a test
fails, it is not immediately clear which change in the batch is responsible for
the failure~\cite{Beheshtian2022}. The current CI system of \product identifies the BIC in the batch
using automatic bisection to aid the bug assignments~\cite{Bach2022}. However, each individual
inspection during the bisection can take up to several hours due to the
compilation, installation, and test execution cost, resulting in severe
bottlenecks in the overall debugging process. The bottleneck can be
particularly problematic if integration or system-level tests fail.

\begin{figure}[t]
  \centerline{\includegraphics[width=0.9\linewidth]{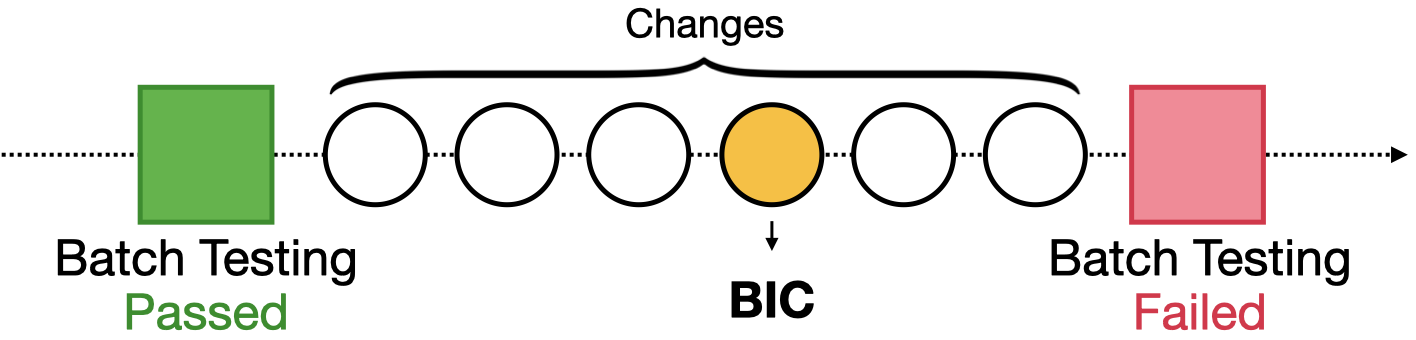}}
  \caption{Simplified batch testing scenario}
  \label{fig:batch_testing}
\end{figure}

This motivates us to see whether \name and its weighted bisection can reduce
the number of bisection iterations. To evaluate the effectiveness of
applying \name, we collect 23 batch testing failures that occurred from July to
August 2022 and their BICs identified by the bisection from the internal CI
logs of \product. Using the data, we first check if \name can find the BIC
inside the batch accurately (Fig.~\ref{fig:batch_testing}). As the test
coverage of \product is regularly and separately updated, instead of being
measured at each of the batch testings, we use the latest line-level coverage
information to calculate the Ochiai scores. Note that we do not need to compute
the Ochiai scores for all lines, but only the lines covered by the failing
tests. When applying \name, depth-based voting decay is not used
($\lambda = 0$) because all candidate commits are submitted on the same day and
have not yet been merged into the main codebase. For the remaining
hyperparameters, we use $\alpha=1$ and $\tau=max$ that performed the best with $\lambda=0$ in RQ1.

\begin{table}[t]
  \centering
  \caption{Evaluation of \name on the 23 Batch Testing Failures of \product}
  \scalebox{1.00}{
    \begin{tabular}{l|r|rrrrr}
        \toprule
     &     \multirow{2}{*}{MRR} & \multicolumn{5}{c}{Accuracy}\\\cmidrule{3-7}
     &           &     @1 &     @2 &     @3 &     @5 &     @10 \\\midrule
     \multirow{2}{*}{\name}     &   \multirow{2}{*}{0.600}   & 10  & 14 & 15 & 20 & 23\\
             &    & (43\%) & (61\%) & (65\%) & (87\%) & (100\%) \\\midrule
    Random & 0.110 & 0 & 0 & 0 & 1 & 17\\\bottomrule
    \end{tabular}}
  \label{tab:industry}
\end{table}

Table~\ref{tab:industry} shows the BIC ranking performance of \name in terms of
MRR and Accuracy@n. While each batch contains 18.48 commits on average, \name
can locate the actual BIC within the top 1 and 5 for 43\% and 87\% of the
failures, respectively. Compared to the random baseline, it achieves 5.5-fold
increase in MRR.
Further, we also report that the weighted bisection can reduce the
bisection iterations for 18 out of 23 cases (78\%), while it increases the cost
in only three cases (13\%). Based on this result, we plan to incorporate weighted
bisection into the CI process of \product, which is expected to save 32\% of
required iterations. Considering that each iteration can take up
to several hours, we expect a significant reduction in the average BIC
identification cost for \product in the long run.

\section{Threats to Validity}
\label{sec:threats}

Threats to internal validity concern factors that can affect how confident we
are about the causal relationship between the treated factors and the effects.
\name relies on widely-adopted open-source tools to establish $\mathsf{Cover}$ and $\mathsf{Evolve}$
relations to ensure the chain of causality between the test failure and BIC
identification. Further evaluation of \name using other code history mining tools such as CodeShovel~\cite{Grund2021} or CodeTracker~\cite{Jodavi2022}, which have been demonstrated to be more accurate than \texttt{git log}, is necessary to further strengthen our claims and will be addressed as future work.
Additionally, other baselines rely on multiple sources of information, such
as bug reports. We choose Defects4J as our benchmark as it provides well-established
links between real-world bug reports and the buggy version, not to mention
human-written bug-revealing test cases that withheld scrutiny from the community.

Threats to external validity concern factors that may affect how well our findings
can be generalised to unseen cases. Our key findings are primarily based on experiments with
the open-source Java programs in Defects4J. Since they are not representative of
the entire population of real-world programs, only further evaluations can strengthen
our claim of generalisation. We tried to support our claim by evaluating \name with
industry-scale software written in C and C++. We do note that \name does not generalise
to bugs that are caused by non-executable files, such as configuration changes, as its
base assumption is that the test failure is caused by a bug in the source code.
We leave extension of \name to bugs caused by non-executable changes as our primary
future work.

Threats to construct validity concern how well the used metrics measure the properties
we aim to evaluate. We adopt two ranking-evaluation metrics, MRR and accuracy@n, to
evaluate \name: both have been widely used in the IR and SE literature. Since they are based on absolute ranks, we do note that the results can be overrated
when the number of ranking candidates is small. To mitigate the threat, we also
present the expected and worst values for the measures as baselines.

\section{Related Work}
\label{sec:related_work}

Locus~\cite{Wen2016} is the first work that proposed to localise the bug at the
software change level. It takes a bug report as an input query and locates the
relevant change hunk based on the token similarities. IR-based techniques, such
as Locus, and \name can complement each other depending on circumstances. When
the failure cannot be reproduced from the bug report, IR-based techniques can
be used instead of \name. However, if the coverage of the failing and passing
tests are available, we can apply \name with SBFL to more precisely rank the
commits without relying on IR.
ChangeLocator~\cite{Wu2017} aims to find a BIC for crashes using the call stack
information. It is a learning-based approach that requires data from fixed
crashes. Unlike ChangeLocator, \name is not limited to crashes and can be
applied to general failures. Orca~\cite{Bhagwan2018} takes symptoms of bugs,
such as an exception message or customer complaints, as an input query and
outputs a ranked list of commits ordered by their relevance to the query. It
uses the TF-IQF~\cite{Yang2008} to compute the relevance scores of files, and
aggregate them to a commit level. Subsequently, it uses machine learning to
predict the risk of candidate commits for breaking ties.
Bug2Commit~\cite{Murali2021} uses multiple features extracted from bug reports
and commits, and aggregates all features by taking the average of their vector
representations. Although Bug2Commit uses an unsupervised learning approach, it
needs the historical data of project-specific bug reports and commits to train
the word embedding model. FBL-BERT~\cite{Ciborowska2022} retrieves the relevant
changeset for the input bug report using a fine-tuned BERT model that can
capture the semantics in the text. It proposes fine-grained changeset encoding methods and accelerates the retrieval by offline indexing~\cite{johnson2019billion}. The major difference between \name and the techniques
mentioned above is that \name does not require any training. Further, \name can
be combined with any code-level FL technique, without being coupled to
specific sources of information, as long as the coverage of failing executions
is available.

The weighted bisection algorithm we propose is similar to FACF (Flaky Aware Culprit Finding)~\cite{Dorward2021}, which formulates the \emph{flake-aware} bisection problem as a Bayesian inference, in that both guide the bisection process based on the probability of commits being a source of test failure. The difference between the two algorithms is that ours uses commit scores from \name to establish the initial probability distribution, while  FACF updates the probability based on the test results during the search taking into account the potential for flakiness. The original work notes that FACF can take into account any prior information about the bug inducing change in the form of an initial probability distribution. Hence, we believe that the commit scores generated by \name can be used as an effective prior distribution for the FACF framework.

There exist studies that are highly relevant to \name despite not being specifically about the BIC identification domain. FaultLocator~\cite{zhang2011localizing} is similar to \name as both use
code-level FL scores to identify suspicious changes. FaultLocator combines spectrum information with the change impact
analysis to precisely identify the failure-inducing \emph{atomic} edits out
of all edits between two versions, whereas \name aims to pinpoint BICs in the
commit history. WhoseFault~\cite{Servant2012} is a method that utilises code-level FL scores and commit history to determine the developer responsible for a bug. While it provides insights into the assignment of bugs, it does not specifically target BIC identification. As a result, it cannot be directly compared with \name in our evaluation, nor can it be integrated with our bisection algorithm. Our belief is that accurately identifying the BIC can also be used to find the developer responsible for fixing the bug, based on the authorship of the changes, in addition to helping developers to understand the context in which the failure occurred.

\section{Conclusion}
\label{sec:conclusion}

This paper proposes \name, a BIC identification technique that is available upon the observation of a failure.
It prunes the BIC search space using failure coverage and the syntactic analysis of commits, and assigns scores
to the remaining commits using the FL scores as well as change histories of code elements. Our experiments with
130 bugs in Defects4J show that \name can effectively identify BICs with an MRR of 0.528, which significantly
outperforms the baselines including state-of-the-art BIC identification techniques. Along with \name, we also
propose the weighted bisection to accelerate the BIC search utilising the commit score information and show
that it can save the search cost in 97.7\% of the studied cases compared to the standard bisection. Finally,
the application of \name to a large-scale industry software \product shows that \name can successfully
reduce the cost of BIC identification in a batch testing CI scenario. Future work includes the actual
deployment of \name to \product as well as expanding the scope of bugs \name can handle.

\section*{Acknowledgment}
We thank three anonymous reviewers whose suggestions helped us improve this paper. Gabin An and Shin Yoo are supported by National Research Foundation of Korea (NRF) Grant (NRF-2020R1A2C1013629), Institute for Information \& communications Technology Promotion grant funded by the Korean government (MSIT) (No.2021-0-01001), and Samsung Electronics (Grant No. IO201210-07969-01).

\bibliographystyle{IEEEtran}
\bibliography{references}

\end{document}